\documentclass[aps,prl,twocolumn,showpacs,groupedaddress]{revtex4}

\usepackage{graphicx}

\begin{document}


\title{Fidelity recovery in chaotic systems and the Debye-Waller factor}


\author{H.-J. St\"ockmann}
\affiliation{Fachbereich Physik der Philipps-Universit\"at Marburg, D-35032 Marburg,
Germany}
\author{R. Sch\"afer}
\affiliation{Fachbereich Physik der Philipps-Universit\"at Marburg, D-35032 Marburg,
Germany}

\date{\today}

\begin{abstract}
Using supersymmetry calculations and random matrix simulations, we
studied the decay of the average of the fidelity amplitude
$f_{\epsilon}(\tau)=\left<\psi(0)|\exp(2\pi\imath
H_{\epsilon}\tau) \exp(-2\pi\imath H_{0}\tau)|\psi(0)\right>$,
where $H_{\epsilon}$ differs from $H_0$ by a slight perturbation
characterized by the parameter $\epsilon$. For strong
perturbations a recovery of $f_{\epsilon}(\tau)$ at the Heisenberg
time $ \tau=1$ is found. It is most pronounced for the Gaussian
symplectic ensemble, and least for the Gaussian orthogonal one.
Using Dyson's Brownian motion model for an eigenvalue crystal, the
recovery is interpreted  in terms of a spectral analogue of the
Debye-Waller factor known from solid state physics, describing the
decrease of X-ray and neutron diffraction peaks with temperature
due to lattice vibrations.

\end{abstract}


\maketitle

\renewcommand {\imath} {{\rm i}}

\newcommand {\xb} {{\bf x}}
\newcommand {\xbd} {{\bf x^\dag}}
\newcommand {\yb} {{\bf y}}
\newcommand {\ybd} {{\bf y^\dag}}
\newcommand {\zn} {{\bf z}_n}
\newcommand {\znd} {{\bf z}_n^{\bf\dag}}
\newcommand {\ree} {R_\epsilon\left(E_1,E_2\right)}
\newcommand {\Tr} {{\rm Tr}}
\newcommand {\rmd} {{\rm d}}
\newcommand{\lla}{\left\langle}
\newcommand{\rra}{\right\rangle}
\newcommand{\rme} {{\rm e}}

\newcommand {\bet}{B}
\newcommand {\gam}{C}

\newcommand {\diag} {{\rm diag}}

The concept of fidelity has been developed by Peres as a tool to characterize the
stability of a quantum-mechanical system against perturbations \cite{per84}. It was
introduced as the squared modulus of the overlap integral of a wave packet with itself
after developing forth and back under the influence of two slightly different
Hamiltonians. Very similar concepts had been applied already in the old spin-echo
experiments of nuclear magnetic resonance half a century ago (see reference \cite{abr61}
for a review). The renewed interest in the topic results from the idea to realize quantum
computers by means of spin systems, where stability against quantum-mechanical
perturbations obviously is of vital importance \cite{fra04}.

Roughly speaking there are three regimes. In the perturbative regime, where the strength
of the perturbation is small compared to the mean level spacing, the decay of the
fidelity is Gaussian. As soon as  the perturbation strength becomes of the order of the
mean level spacing, exponential decay starts to dominate, with a decay constant obtained
from Fermi's golden rule \cite{cer02, jac01b}. For very strong perturbations the decay
becomes independent of the strength of the perturbation. Here, in the Loschmidt regime,
the decay is still exponential, but now the decay constant is given by the classical
Lyapunov exponent \cite{Jal01}. Exactly such a behaviour had been observed experimentally
in a spin-echo experiment on isolated spins coupled weakly to a bath of surrounding spins
\cite{pas95}.

Gorin et al. \cite{gor04} calculated the  fidelity decay within  random matrix theory in
the regime of small perturbations and could correctly describe the change from Gaussian
to exponential behaviour with increasing perturbation strength. The Lyapunov regime is
non-universal and thus not accessible in a random matrix model.

Intuitively, one would expect that in chaotic systems the fidelity decay is stronger than
in integrable ones. The opposite is true. Prosen et al. \cite{pro02b} showed that a
chaotic system is much more {\em fid\`ele} than a regular one, and suggested to use
chaotic systems in quantum computing to suppress chaos.

It will be shown here that this is only part of the truth, and
that for chaotic systems there is even a partial recovery of the
fidelity at the Heisenberg time. This work extends the results by
Gorin et al. \cite{gor04} to the regime of strong perturbations
using supersymmetry techniques. It is stressed that our results
are generic and not restricted to random matrix systems. In a
spin-chain model, e.\,g., the fidelity recovery has been observed
recently as well \cite{pin}.

Using the Brownian-motion model for the eigenvalues of random matrices introduced by
Dyson many years ago \cite{dys62a}, it will be shown that this behaviour has its direct
analogue in the Debye-Waller factor of solid state physics. We shall sketch the
calculation for the GUE only, and will just cite the result for the GOE. More details
will be presented in a forthcoming paper \cite{stoe04b}.

Let us start with an unperturbed Hamiltonian $H_0$, which is
turned by a small perturbation into
\begin{equation}\label{00c}
  H_\phi=H_0\cos\phi+ H_1\sin\phi\,
\end{equation}
It is assumed that both $H_0$ and $H_1$ have a mean level spacing
of one, i.\,e. the variance of the off-diagonal elements is given
by $N/\pi^2$ \cite{meh91}. This particular choice of the
perturbation guarantees that the mean density of states does not
change with $\epsilon$. It will be come clear below that $\phi$
has to be of ${\cal O}(1/\sqrt{N})$ to allow a well-defined limit
$N\to\infty$. We therefore introduce $\epsilon=4N\phi^2$ as the
perturbation parameter. It follows for the fidelity amplitude
\begin{equation}\label{00}
  f_{\epsilon}(\tau)=\left<\psi(0)\left|e^{2\pi\imath (cH_0+ sH_1)\tau}
  e^{-2\pi\imath H_0\tau}\right|\psi(0)\right>\,,
\end{equation}
where $\psi(0)$ is the wave function at the beginning, and
$c=\cos\phi$, $s=\sin\phi$.  The squared modulus of
$f_{\epsilon}(\tau)$ yields the fidelity $F_{\epsilon}(\tau)$,
originally introduced by Peres \cite{per84}. The calculation of
the average of $F_{\epsilon}(\tau)$, however, is technically more
involved and therefore not considered here.

For the present work the paper of Gorin et al \cite{gor04} is of
particular relevance. The author considered a slightly different
parameter dependence
\begin{equation}\label{000}
    H_\lambda=H_0+\lambda V\,,
\end{equation}
where the variance of the off-diagonal elements of $V$ was assumed
to be one. Comparison with equation (\ref{00c}) shows that the
respective perturbation parameters are related via
$\lambda=\sqrt{\epsilon}/(2\pi)$.

In the paper by Gorin et al. \cite{gor04} the Gaussian average of
the fidelity amplitude was calculated in the regime of small
perturbation strengths, correct up to ${\cal O}(\epsilon)$,
\begin{equation}\label{00a}
  \langle f_{\epsilon}(\tau)\rangle\sim e^{- \epsilon \, C(\tau)}\,.
\end{equation}
where $C(\tau)$ is given by
\begin{equation}\label{00aa}
    C(\tau)=\frac{\tau^2}{\beta}+\frac{\tau}{2}-\int_0^\tau \int_0^t
b_2(t^\prime) \rmd t^\prime \rmd t \,,
\end{equation}
and $b_2(\tau)$ is the spectral form factor. $\beta$ is Dyson's
universality index, i.\,e. $\beta=1$ for the Gaussian orthogonal
ensemble (GOE), $\beta=2$ for the Gaussian unitary ensemble (GUE),
and $\beta=4$ for the Gaussian symplectic ensemble (GSE). Equation
(\ref{00a}) describes correctly the change from Gaussian to
exponential decay with increasing perturbation strength. Equation
(\ref{00a}) was the motivation for the introduction of $\epsilon$
as the perturbation parameter.

In chaotic systems the average of the fidelity amplitude over the initial state $\psi(0)$
reduces to a trace,
\begin{equation}\label{01}
  \langle f_\epsilon(\tau) \rangle=\frac{1}{N}\left<{\rm Tr}\left[e^{2\pi\imath (cH_0+ sH_1)\tau}
  e^{-2\pi\imath H_0\tau}\right]\right>\,.
\end{equation}
$\langle f_\epsilon(\tau) \rangle$ may be written as a Fourier
transform,
\begin{equation}\label{02}
  \langle f_\epsilon(\tau) \rangle=\int dE_1\,dE_2
  e^{2\pi\imath\left(E_1-E_2\right)\tau}\,\ree\,
\end{equation}
where
\begin{eqnarray}\label{03}
  \lefteqn{\ree\sim}&&\nonumber\\&&\frac{1}{N}\left<{\Tr}\left(
  \frac{1}{E_{1-}-cH_0-sH_1}\,\frac{1}{E_{2+}-H_0}\right)\right>\,,\nonumber\\
\end{eqnarray}
with $E_\pm=E\pm\imath\eta$.  Using standard supersymmetry techniques \cite{ver85a}, this
can be written as
\begin{eqnarray}\label{04}
  \lefteqn{\ree\sim}&&\nonumber\\&&\frac{1}{N}
  \int d[x]\,d[y]\,\sum\limits_{n,m}
  (x_n^*x_m-\xi_n^*\xi_m)(y_m^*y_n-\eta_m^*\eta_n)\nonumber
  \\&&\times e^{-\imath\left[\xbd E_1\xb-\ybd
  E_2\yb\right]}\nonumber\\&&\times
  \left<e^{\imath \left[c\xbd H_0\xb-\ybd H_0 \yb\right]}\right>\,
  \left<e^{\imath s\xbd H_1\xb}\right>\,,
\end{eqnarray}
where $\xb=\left(x_1,\xi_1\dots x_N,\xi_N\right)^T$, $\yb=\left(y_1,\eta_1\dots
y_N,\eta_N\right)^T$, and
\begin{equation}\label{05}
d[x]=\prod_n dx_n\,dx_n^*\,d\xi_n\,d\xi_n^*\,,\quad d[y]=\prod_n
dy_n\,dy_n^*\,d\eta_n\,d\eta_n^*\,.\nonumber
\end{equation}

We adopt the usual convention using latin letters for commuting, and greek ones for
anticommuting variables, respectively. Now the Gaussian average over the matrix elements
of $H_0$ and $H_1$ can be performed elementary.

The subsequent steps (Hubbard-Stratonovich transformation, integration over the $x$, $y$
variables, saddle point integration etc.) are essentially the same ones as for the
calculation of the spectral form factor (see e.\,g. chapter 10 of reference \cite{haa01b}
for the GUE case). Details  can be found in Ref. \onlinecite{stoe04b}. We then arrive
at the result
\begin{eqnarray}\label{37}
  \ree&\sim&\frac{\rho^2}{N}
  \int\limits_0^\infty dx\int\limits_0^1 dy
  \frac{x+y+x^2-y^2}
  {(x+y)^2}\nonumber\\&&\times
  e^{-2\pi\imath\rho E(x+y)}
  e^{-\frac{\epsilon}{2}\rho^2(x+y)(1+x-y)}\,,
\end{eqnarray}
where $\rho=\sqrt{1-[\pi\bar{E}/(2N)]^2}$ is the mean density of states, and
$\bar{E}=(E_1+E_2)/2$, $E=E_1-E_2$. Inserting this result into equation~(\ref{02}), and
introducing $\bar{E}$ and $E$ as new integration variables, we get, fixing the constant
of proportionality by the condition $f_\epsilon(0)=1$,
\begin{eqnarray}\label{38}
  \langle f_\epsilon(\tau) \rangle&=&\frac{1}{N}\int d\bar{E}\rho^2
  \int\limits_0^\infty dx\int\limits_0^1 dy
  \frac{1+x-y}
  {x+y}\nonumber\\&&\times
  \delta [\tau-\rho(x+y)]
  e^{-\frac{\epsilon}{2}\rho^2(x+y)(1+x-y)}\,.
  \end{eqnarray}
The $\bar{E}$ integration is nothing but an energy average. In the
limit $N\to\infty$ only the band centre contributes where  $\rho$
takes the constant value one. We may then discard this average and
obtain
\begin{equation}\label{40}
  \langle f_\epsilon(\tau) \rangle=\frac{1}{\tau}\int_0^{{\rm Min}(\tau,1)}dy
  (1+\tau-2y)e^{-\frac{\epsilon}{2}\tau(1+\tau-2y)}\,.
\end{equation}
The integral is easily performed with the result
\begin{equation}\label{41}
  \langle f_\epsilon(\tau) \rangle=\left\{\begin{array}{ll}
    e^{-\frac{1}{2}\epsilon\tau}\left[s\left(\frac{1}{2}\epsilon\tau^2\right)
    -\tau s'\left(\frac{1}{2}\epsilon\tau^2\right)\right]\,,\qquad &\tau\le1 \\
    e^{-\frac{1}{2}\epsilon\tau^2}\left[s\left(\frac{1}{2}\epsilon\tau\right)
    -\frac{1}{\tau}s'\left(\frac{1}{2}\epsilon\tau\right)\right]\,,&\tau>
    1
  \end{array}\right.\,,
\end{equation}
where
\begin{equation}\label{42}
  s(x)=\frac{\sinh(x)}{x}\,.
\end{equation}

We have thus obtained an analytic expression for the GUE average of the fidelity
amplitude for arbitrary perturbation strengths. In the limit of small perturbations it is
in complete accordance with the result obtained by Gorin et al. \cite{gor04}.

The calculation for the GOE is done in exactly the same way. It is technically much more
involved, but fortunately most of the work for this case has already been done  by
Verbaarschot, Weidenm\"uller, and Zirnbauer in their disseminating work \cite{ver85a}. In
this way we get for the GOE average of the fidelity amplitude
\begin{eqnarray}\label{57}
  \langle f_\epsilon(\tau) \rangle&=&2\int\limits_{{\rm Max}(0,\tau-1)}^\tau du\int\limits_0^u
  \frac{v\,dv}{\sqrt{[u^2-v^2][(u+1)^2-v^2]}}\nonumber\\
  &&\times\frac{(\tau-u)(1-\tau+u)}{(v^2-\tau^2)^2}\\
  &&\times
  [(2u+1)\tau-\tau^2+v^2]e^{-\frac{1}{2}\epsilon[(2u+1)\tau-\tau^2+v^2]} \, . \nonumber
\end{eqnarray}

To affirm the analytical findings, random matrix simulations were performed for all
Gaussian ensembles including the  symplectic one, which has not been treated
analytically. The average was taken over up to 8000 random matrices of rank $N=500$ for
$H_0$, and for each of them over 50 random matrices for $H_1$.

\begin{figure}
\begin{center}
  \includegraphics[width=0.44\textwidth]{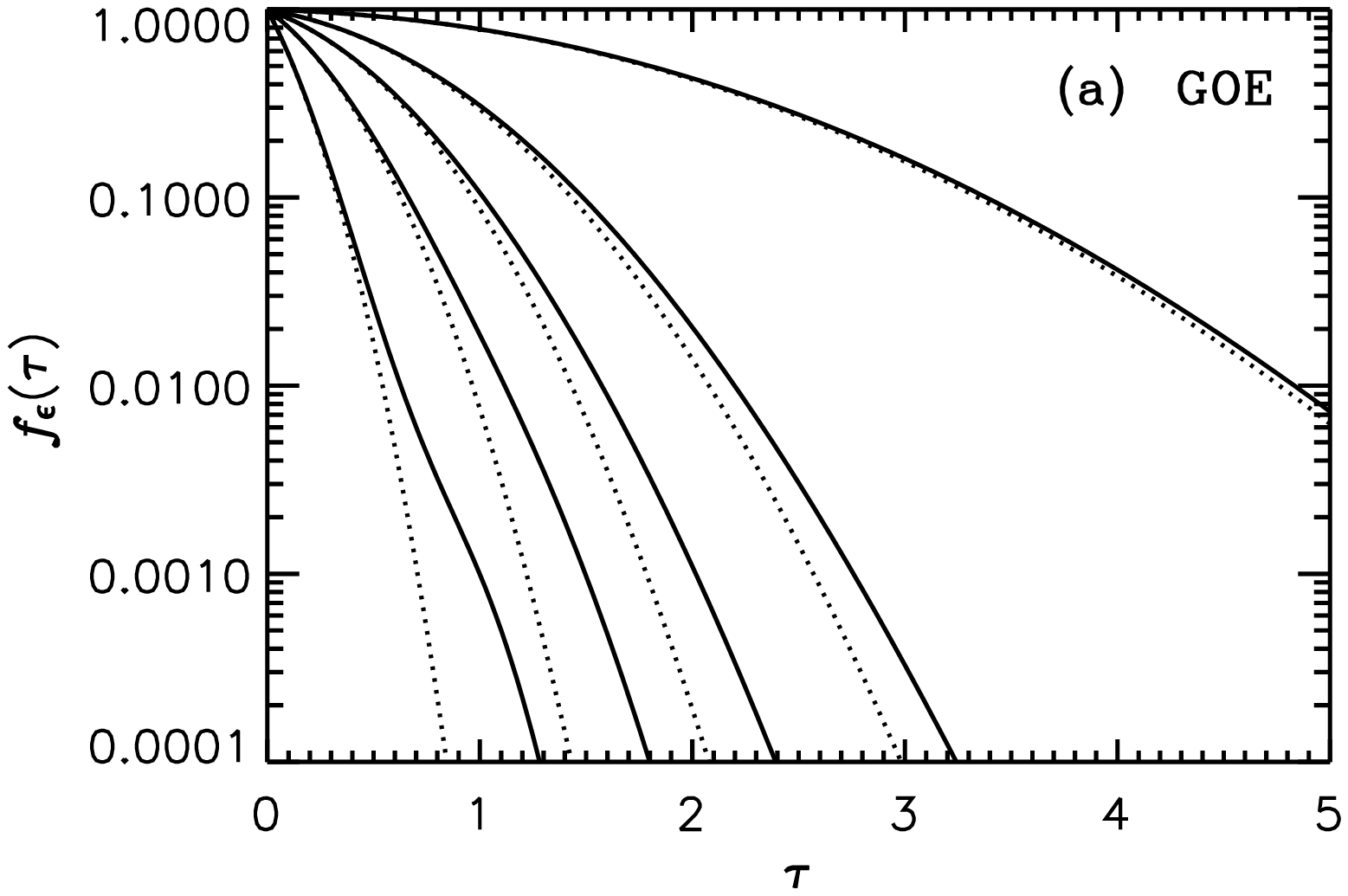}\\[-4.5ex]
 \includegraphics[width=0.44\textwidth]{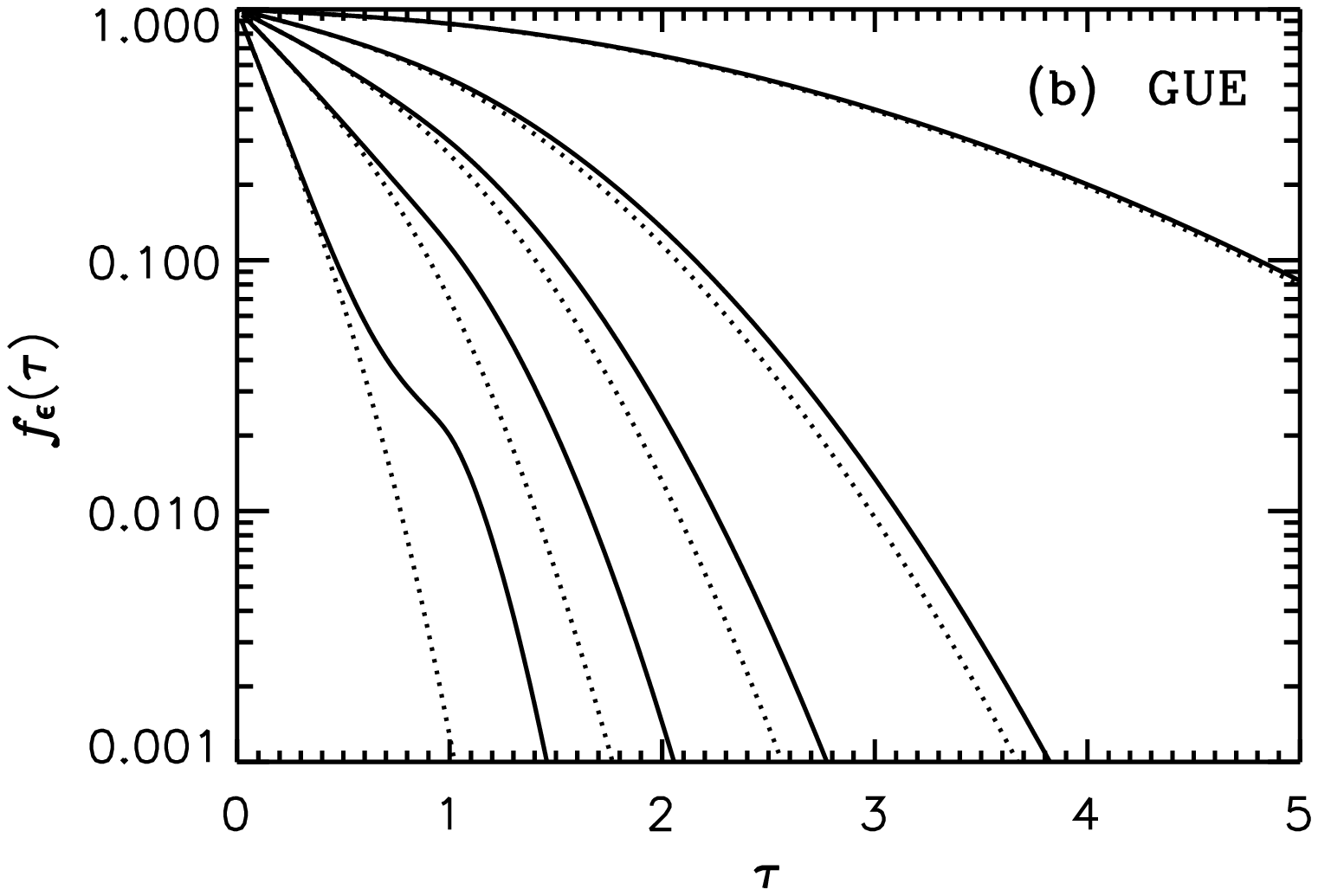}\\[-4.5ex]
 \includegraphics[width=0.44\textwidth]{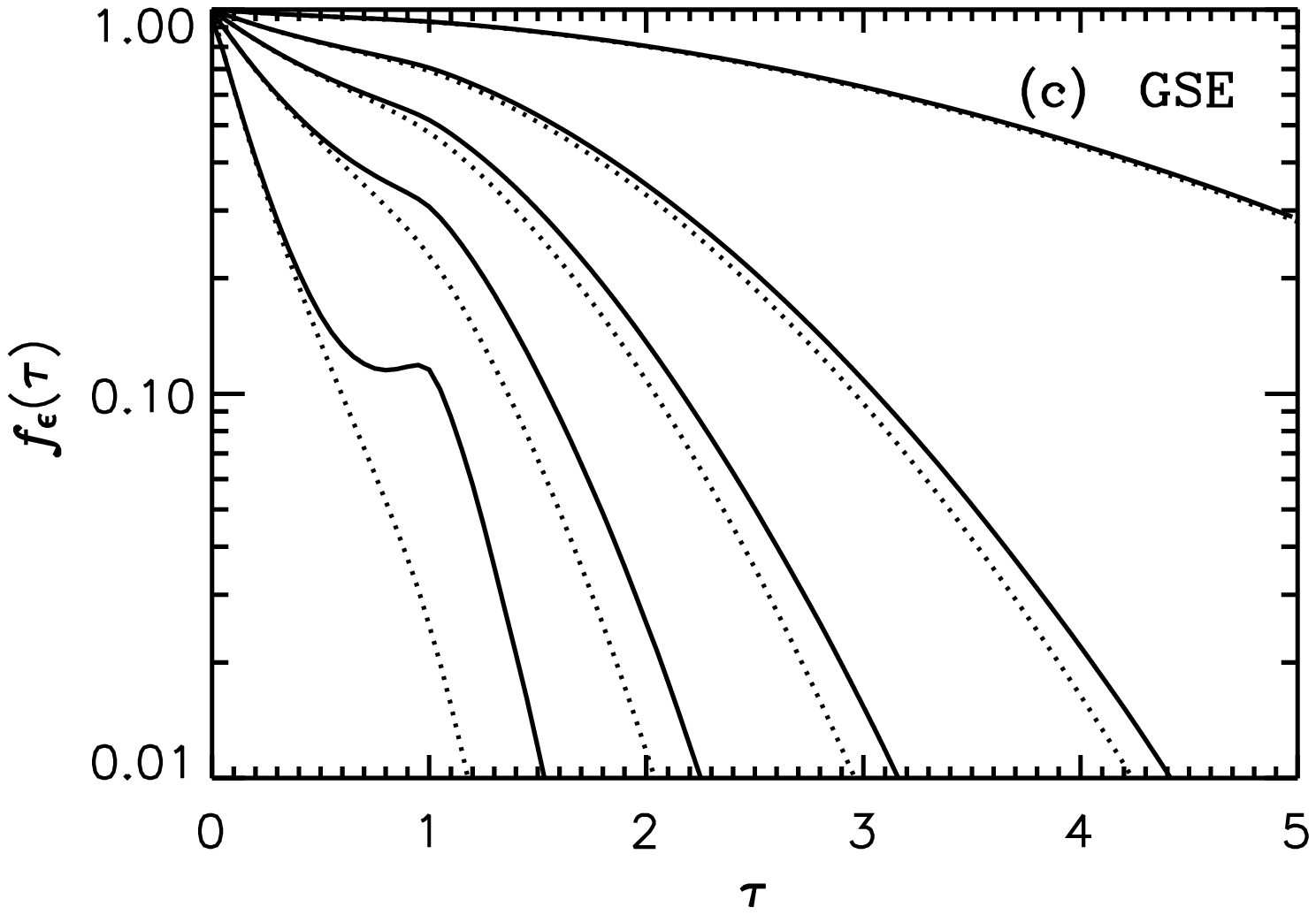}
  \caption{Average of the fidelity amplitude $\langle f_\epsilon(\tau) \rangle$ for the GOE (a), the GUE (b),
  and the GSE (c) for $\epsilon=0.2$, 1, 2, 4 and 10. $\tau$ is given in units of the Heisenberg time.
   For the GOE and the GUE the  solid lines show the results of the analytical calculation,
   for the GSE of the numerical simulation. The dotted lines
   correspond to the linear response approximation (\ref{00a}). The numerical results are in agreement with the analytical
   results within the limits of the line strength.
  }\label{fig:vgl_gue_goe}
\end{center}
\end{figure}

The results are shown in Figure \ref{fig:vgl_gue_goe}. For the GOE and the GUE the
numerical simulations are in perfect agreement with the analytical results for all
$\epsilon$ values shown. For comparison,  the fidelity amplitudes in the linear response
approximation (\ref{00a}) are presented as well. For small perturbation strengths or
small values of $\tau$ the linear response result is a good approximation, but the limits
of its validity are also clearly illustrated.

\begin{figure}
\begin{center}
  \includegraphics[width=0.44\textwidth]{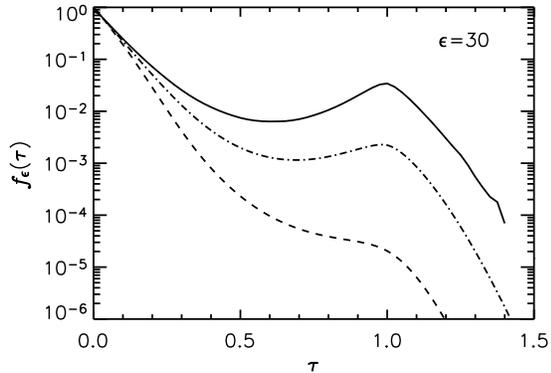}
  \caption{Average of the fidelity amplitude $\langle f_\epsilon(\tau) \rangle$ for
  $\epsilon=30$.
   The dashed and dashed-dotted lines show the analytical results for the GOE and the GUE,
   respectively. The solid line corresponds to the numerical simulations for the GSE, reliable up to $10^{-3}$.
  }\label{fig:vgl_ous}
\end{center}
\end{figure}

For small perturbation strengths $\epsilon$ the decay of the fidelity is predominantly
Gaussian which changes into a behaviour showing a cross-over from an exponential to a
Gaussian decay at $\epsilon\approx 1$, in accordance with literature. The most
conspicuous result of the present letter, however, is the partial recovery of the
fidelity at the Heisenberg time $\tau=1$ which has not been reported previously to the
best of our knowledge. This recovery is most pronounced for the GSE and least for the
GOE. This is illustrated in Figure \ref{fig:vgl_ous} showing a direct comparison of the
fidelity amplitudes of the Gaussian ensembles for $\epsilon=30$. Preliminary
simulations showed that the fidelity recovery is observed not only in the fidelity
amplitude, but in the fidelity as well.

What is the origin of the surprising recovery? We believe that there is a simple
intuitive explanation in terms of Dyson's Brownian motion model \cite{dys62a}. Since the
mean density of states is kept constant during the parameter change, the eigenvalues of
$H_\phi$  may be written as $E_k^{(\phi)}=k+\delta_k^{(\phi)}$, where $\delta_k^{(\phi)}$
fluctuates about zero.  For strong perturbations the eigenvectors of the perturbed and
unperturbed system are uncorrelated, and we get
\begin{equation}\label{61}
  \langle f_\epsilon(\tau) \rangle=\frac{1}{N}\sum_{kl}\left<|R_{lk}|^2\right>e^{2\pi\imath\tau(k-l)}W\,,
\end{equation}
where the $R_{kl}$ are obtained elementary from the eigenvectors, and $W$ is given by
\begin{equation}\label{62}
  W=\left<e^{2\pi\imath\tau\left(\delta_k^{(\phi)}-\delta_l^{(0)}\right)}\right>
  \approx e^{-(2\pi\tau)^2\left<\delta^2\right>}\,.
\end{equation}

In the second step a Gaussian approximation was applied. Within the framework of the
Brownian motion model $\left<\delta^2\right>$ is interpreted as the mean squared
displacement of an eigenvalue from its equilibrium position. It is proportional to
``temperature'' $T$, which is just the reciprocal universality factor $\beta$, whence
follows
\begin{equation}\label{63}
  W=e^{-\alpha\tau^2T}
\end{equation}
with some constant $\alpha$. It follows from equation (\ref{61}) that there is a revival
of the fidelity at the Heisenberg time $\tau=1$ decreasing with ``temperature''
proportional to $e^{-\alpha T}$. This is exactly the behaviour illustrated in Figure
\ref{fig:vgl_ous}.

There is a perfect analogy to the  temperature dependence of  X-ray and neutron
diffraction patterns in solid state physics.  Caused by lattice vibrations the
intensities of the diffraction maxima decrease with temperature with a dependence
described by the Debye-Waller factor
\begin{equation}\label{64}
  W_{\rm DW}=e^{-\beta g^2T}\,,
\end{equation}
where $\beta$ is another constant, and $g$ is the modulus of the reciprocal lattice
vector characterizing the reflex (see e.\,g. appendix A of reference \cite{kit99}). This
is our justification to call $W$ a spectral Debye-Waller factor.

One may argue that due to equation (\ref{61}) there should be revivals for all integer
multiples of the Heisenberg time, which are not observed. This  can be understood by
considering the analogy between the spectral form factor
\begin{equation}\label{65}
  K(\tau)=\frac{1}{N}\sum\limits_{n,m}
  e^{2\pi\imath\left(E_n-E_m\right)\tau}
\end{equation}
and the structure factor in condensed matter,
\begin{equation}\label{66}
  S(\vec{k})=\frac{1}{N}\sum\limits_{n,m}
  e^{2\pi\imath\vec{k} \, \cdot \left(\vec{R}_n-\vec{R}_m\right)}
\end{equation}
where the $\vec{R}_n$ are the positions of the atoms, and
$\vec{k}$ is a point in the reciprocal lattice. In liquids and
glasses these peaks are smeared out, and the structure factor
depends only on $k=|\vec{k}|$. $S(k)$ starts at zero for $k=0$,
climbs up to two to three at the $k$ value corresponding to the
reciprocal atomic distance, oscillates about one with decreasing
amplitude for larger $k$ values, and approaches one in the limit
$k\to\infty$ \cite{zim79}. There is a striking similarity with the
spectral form factor $K(\tau)$ of the GSE, the eigenvalue
``crystal'' with the lowest available temperature. $K(\tau)$, too,
starts at zero for $\tau=0$, has a logarithmic singularity at the
Heisenberg time $\tau=1$, and approaches one for $\tau\to\infty$.
The oscillations found in $S(k)$, however, are absent in
$K(\tau)$.
Fidelity amplitude and spectral form factor thus do not show any
structure at multiple integers of the Heisenberg time, in contrast
to the corresponding condensed-matter quantities. The explanation
is straightforward: Pauli's principle prevents the atoms from
approaching too closely, whereas in the eigenvalue ``crystal''
such a lower limit does not exit.

If it is true that the universal spectral properties for all
chaotic systems are described correctly by random matrix theory
\cite{boh84b}, and there is overwhelming evidence for this fact,
then the exact expressions for the fidelity amplitude decay
derived in this paper are universal and hold for all chaotic
systems.  But this means that the fidelity decay in this regime
reflects the phase space properties of the system, but not its
stability against perturbations as originally claimed by Peres
\cite{per84}. Only in the regime of the Loschmidt echo the system
stability shows up.

\begin{acknowledgments}
This paper has profited a lot from numerous discussions on the subject of fidelity with
Thomas Seligman, Cuernavaca, Mexico, Thomas Gorin, Freiburg, Germany and Toma{\v z}
Prosen, Ljubljana, Slovenia. Thomas Guhr, Lund, Sweden is thanked for discussions of the
supersymmetry aspects of this paper. The work was supported by the Deutsche
Forschungsgemeinschaft.
\end{acknowledgments}


\end{document}